\pdfoutput=1

\documentclass[final,5p,times,twocolumn]{elsarticle}

\usepackage{graphicx}

\newcommand {\Vector}[1]{\mbox{\boldmath $#1$}}
\newcommand {\Matrix}[1]{\mbox{\boldmath $#1$}}

\newcommand {\Set}[1]{\mathbb{#1}}

\usepackage{colortbl} 
\usepackage{multirow} 
\usepackage{amsmath}

\usepackage{amssymb}

\journal{Speech Communication}

\begin{document}

\begin{frontmatter}

\title{Model architectures to extrapolate emotional expressions in DNN-based text-to-speech} 


\author[label1]{Katsuki Inoue}
\address[label1]{Graduate school of Interdisciplinary Science and Engineering in Health Systems, Okayama University, Japan}

\author[label1]{Sunao Hara}

\author[label1]{Masanobu Abe}

\author[label3]{Nobukatsu Hojo}
\address[label3]{NTT Corporation, Japan}

\author[label3]{Yusuke Ijima}
%
%
%
%

\begin{abstract}

This paper proposes architectures that facilitate the extrapolation of emotional expressions in deep neural network (DNN)-based text-to-speech (TTS).
In this study, the meaning of ``extrapolate emotional expressions'' is to borrow emotional expressions from others, and the collection of emotional speech uttered by target speakers is unnecessary. 
Although a DNN has potential power to construct DNN-based TTS with emotional expressions and some DNN-based TTS systems have demonstrated satisfactory performances in the expression of the diversity of human speech, it is necessary and troublesome to collect emotional speech uttered by target speakers. 
To solve this issue, we propose architectures to separately train the speaker feature and the emotional feature and to synthesize speech with any combined quality of speakers and emotions.
The architectures are parallel model (PM), serial model (SM), auxiliary input model (AIM), and hybrid models (PM\&AIM and SM\&AIM).
These models are trained through emotional speech uttered by few speakers and neutral speech uttered by many speakers.
Objective evaluations demonstrate that the performances in the open-emotion test provide insufficient information.
They make a comparison with those in the closed-emotion test, but each speaker has their own manner of expressing emotion. 
However, subjective evaluation results indicate that the proposed models could convey emotional information to some extent. 
Notably, the PM can correctly convey sad and joyful emotions at a rate of ${>}60$\%. 

\end{abstract}


\end{frontmatter}


\section{Introduction}\label{sec:Intro}

This paper proposes architectures that facilitate the extrapolation of emotional expressions in deep neural network (DNN)-based text-to-speech (TTS).
Text-to-speech (TTS), is a technology that generates speech from text.
A variety of TTS methods have been proposed to generate natural, intelligible, and human-like speech.
Recently, deep neural network (DNN)-based TTS has been intensively investigated, and the results demonstrated that DNN-based TTS can outperform hidden Markov model (HMM)-based TTS in the quality and naturalness of synthesized speech \cite{bib-DNN-TTS-01,bib-DNN-TTS-02,bib-DNN-TTS-03,bib-HMMvsDNN-01}.
First, a feed-forward neural network (FFNN) was proposed as a replacement for decision-tree approaches in HMM-based TTS \cite{bib-DNN-TTS-01}.
Subsequently, long short-term memory (LSTM)-based recurrent neural network (RNN) has been adopted and provided better naturalness and prosody because of their capability to model the long-term dependencies of speech \cite{bib-LSTM-RNN-TTS-01}.

In addition to the quality and naturalness, DNN-based TTS has advantages in capabilities for controlling voice aspects.
For example, to control speaker identity, several multi-speaker models have been proposed \cite{bib-DNN-MultiSpeaker-01,bib-DNN-MultiSpeaker-02,bib-DNN-MultiSpeaker-03_IEICE}.
To control speaker changes, a method using auxiliary vectors of the voice's gender, age, and identity was proposed \cite{bib-DNN-MultiAgeGender-01}. 
Additionally, a multi-language and multi-speaker model was built by sharing data across languages and speakers \cite{bib-DNN-MultiLangSpeaker-01}.

In terms of voice aspects, emotional expression is one of the more important features.
Notably, DNN-based TTS was additionally proposed to synthesize speech with emotions.
For example, in the simplest approach, usage of the emotional one-hot vector was proposed by \cite{bib-DNN-MultiEmotion-01}.
To control emotional strength, a method that used an auxiliary vector based on listener perception was proposed by \cite{bib-DNN-MultiEmotion-02}.
By using a speaker adaptation method, Yang et. al\ \cite{bib-DNN-MultiEmotion-03} proposed a method that generated emotional speech from a small amount of emotional speech training data.
In all aforementioned methods, the target speaker's emotional speech is necessary for training.
However, in general, it is difficult for individuals to utter speech with a specific emotion and to continue speaking with the emotion for a few hours.
In conclusion, recording the target speaker's emotional speech is a bottleneck in constructing DNN-based TTS that can synthesize emotional speech with a particular speaker's voice quality.


To overcome the problem, one possible approach is extrapolation.
Emotional expression models are trained using speech uttered by a particular person, and the models are applied to another person to generate emotional speech with the person's voice quality. 
In other words, a collection of emotional speech uttered by a target speaker is not required, and emotional expression is generated using models trained by the emotional speech of another person. 
In summary, the meaning of extrapolation is to borrow emotional models from another
individual. 
Based on this approach, several methods have been proposed, for example, in HMM-based TTS, methods that can generate emotional speech in extrapolation conditions.
Kanagawa et al.\ suggested generating speaker-independent transformation matrices using pairs of neutral and target-style speech, and applying these matrices to a neutral-style model of a new speaker \cite{bib-HMM-MultiEmotion-open-01}.
Similarly, Trueba et al.\ \cite{bib-HMM-MultiEmotion-open-02} proposed to extrapolate the expressiveness of proven speaking-style models from speakers who utter speech in a neutral speaking style.
The proposal included using a constrained structural maximum a posteriori linear regression (CSMAPLR) algorithm \cite{bib-CSMAPLR}.
Ohtani et al.\ proposed an emotion additive model to extrapolate emotional expression for a neutral voice \cite{bib-HMM-MultiEmotion-open-03}.
All the aforementioned methods above suggest that the extrapolation of emotional expressions is possible by separately modeling the emotional expressions and the speaker identities.

Based on the extrapolation approach, we propose a novel DNN-based TTS that can synthesize emotional speech.
The biggest advantage of the proposed algorithm is that we can synthesize several types of emotional speech with the voice qualities of the multiple speakers.
This works even if the target speaker’s emotional speech is not included in training data.
A key idea is to explicitly control the speaker factor and the emotional factor, motivated by the success in the multi-speaker model \cite{bib-DNN-MultiSpeaker-01,bib-DNN-MultiSpeaker-02,bib-DNN-MultiSpeaker-03_IEICE,bib-DNN-MultiAgeGender-01,bib-DNN-MultiLangSpeaker-01} and multi-emotional model \cite{bib-DNN-MultiEmotion-01,bib-DNN-MultiEmotion-02,bib-DNN-MultiEmotion-03}. 
Once the factors are trained, by independently controlling the factors, we can synthesize speech with any combination of a speaker and an emotion.
As training data, we have emotional speech uttered by few speakers, including a neutral speaking style, and have only neutral speech uttered by many speakers. 
The speaker factor must be trained using the neutral speech uttered by each speaker, and the emotional factor must be trained using the speech uttered by few speakers. 
To achieve the purpose, we examine five types of DNN architectures: parallel model (PM), serial model (SM), auxiliary input model (AIM), and the hybrid models (PM\&AIM and SM\&AIM).
The PM deals with emotional factors and speaker factors in parallel on the output layer.
Also, the SM deals with two factors in serial order on the last hidden layer and output layer.
The AIM deals with the two factors by using auxiliary one-hot vectors.
Differing from those simple models, the hybrid models are composed two potential pairs: PM and AIM, or SM and AIM.
In \cite{inoue2017investigation}, we reported the extrapolation of emotional expressions in acoustic feature modeling, and evaluated the performance of synthesized speech uttered by only female speakers.
Additionally, in this paper, we investigate the extrapolation of emotional expressions in phoneme duration modeling, and evaluate the performance of synthesized speech uttered by both males and females.
This paper is organized as follows.
In Section $2$, we provide an overview of DNN-based TTS and introduce expansions to control multiple voice aspects.
In Section $3$, we describe the proposed DNN architectures.
In Section $4$, we explain objective and subjective evaluation.
In Section $5$, we present our conclusions and suggestions for
further research.

\section{DNN-based TTS}\label{sec:DNN-TTS}


DNN-based TTS is a method of speech synthesis that uses the DNN to map linguistic features to acoustic features.
A DNN-based TTS system comprises of text analysis, a phoneme duration model, an acoustic model, and waveform synthesis.


The simplest DNN that generates output vector $\Vector{y}$ from input vector $\Vector{x}$ is expressed by following a recursive formula.
		\begin{equation}
		\begin{split}
		&\Vector{h}^{(\ell)} = f^{(\ell)}( \Matrix{W}^{(\ell)} \Vector{h}^{(\ell-1)} + \Vector{b}^{(\ell)} )
		\\
		&\mathrm{where\ \ } 1 \leq \ell \leq L, \  \Vector{h}^{(0)}= \Vector{x}, \ \Vector{h}^{(L)}= \Vector{y}.
		\end{split}
		\label{eq:DNN1}
		\end{equation}
$\Vector{h}^{(\ell-1)} \in \Set{R}^{d_{\ell-1} \times 1}$ is the $d_{\ell-1}$ dimensional output vector of $(\ell-1)$-th layer, and $\Vector{h}^{(\ell)} \in \Set{R}^{d_{\ell} \times 1}$ is the $d_{\ell}$ dimensional output vector of $\ell$-th layer.
Additionally, $\Matrix{W}^{(\ell)} \in \Set{R}^{d_{\ell} \times d_{\ell-1}}$ and $\Vector{b}^{(\ell)} \in \Set{R}^{d_{\ell} \times 1}$ are the weight matrix, with bias from the $(\ell-1)$-th hidden layer to the $\ell$-th hidden layer,  $f^{(\ell)}(\cdot)$ is the activation function on the $\ell$-th hidden layer, and $L$-th layer is the output layer.

To control the voice aspects, for example, the speaker identity, speaking style, and emotional expression, the DNN architecture was expanded in two ways: input and output.
Expansion of input is a common way to control the voice aspects and is known as feature-embedding.
Also, expansion of output is a newly proposed way in our research~\cite{inoue2017investigation} and inspired by multitask learning DNN \cite{bib-DNN-TTS-03}.

\subsection{Expansion of input} 

The input of the $\ell$-th layer in eq.\ref{eq:DNN1} $\Vector{h}^{(\ell-1)}$ can be expanded as follows,
		\begin{align}
		\label{eq:DNN-IN1}
		\Vector{h}^{(\ell)} &= f^{(\ell)}( \Matrix{W}^{(\ell)}_{a} \Vector{h}^{(\ell-1)}_{a} + \Vector{b}^{(\ell)} ) \\
		\nonumber \\
		\Vector{h}^{(\ell-1)}_{a} &= 
		\left[
		\begin{array}{l}
		\label{eq:DNN-IN2}
		\Vector{h}^{(\ell-1)}\\
		\Vector{v}^{(\ell-1)}_{a}
		\end{array}
		\right]\\
		\label{eq:DNN-IN3}
		\Matrix{W}^{(\ell)}_{a} &= \left[ \Matrix{W}^{(\ell)} \quad \Matrix{W}_{a} \right]
		\end{align}
The input vector $\Vector{h}^{(\ell-1)}_{a} \in \Set{R}^{(d_{\ell-1}+d_{a}) \times 1}$ comprises the input to the $\ell$-th layer $\Vector{h}^{(\ell-1)}$ and the auxiliary vector $\Vector{v}^{(\ell-1)}_{a} \in \Set{R}^{d_{a} \times 1}$.
$\Matrix{W}^{(\ell)}_{a} \in \Set{R}^{d_{\ell} \times (d_{\ell-1}+d_{a})}$ is the weight matrix in the $\ell$-th layer.
The effects caused by the auxiliary vector are spread throughout the entire model, which results in controlling the factors as a black box.


As the auxiliary vector, An et al. \cite{bib-DNN-MultiEmotion-01} applied a one-hot vector that indicated emotions to all layers ($1 \leq \ell \leq L$) for multi-emotional modeling.
Additionally, Wu et al. \cite{bib-DNN-MultiSpeaker-02} applied an i-vector \cite{bib-i-vector} to the first hidden layer ($\ell=1$) for multi-speaker modeling, and Hojo et al. \cite{bib-DNN-MultiSpeaker-03_IEICE} applied a one-hot vector to all layers ($1 \leq \ell \leq L$) for multi-speaker modeling.

		\begin{figure*}[bt]
			\begin{center}
			\includegraphics[width=1.0\textwidth]{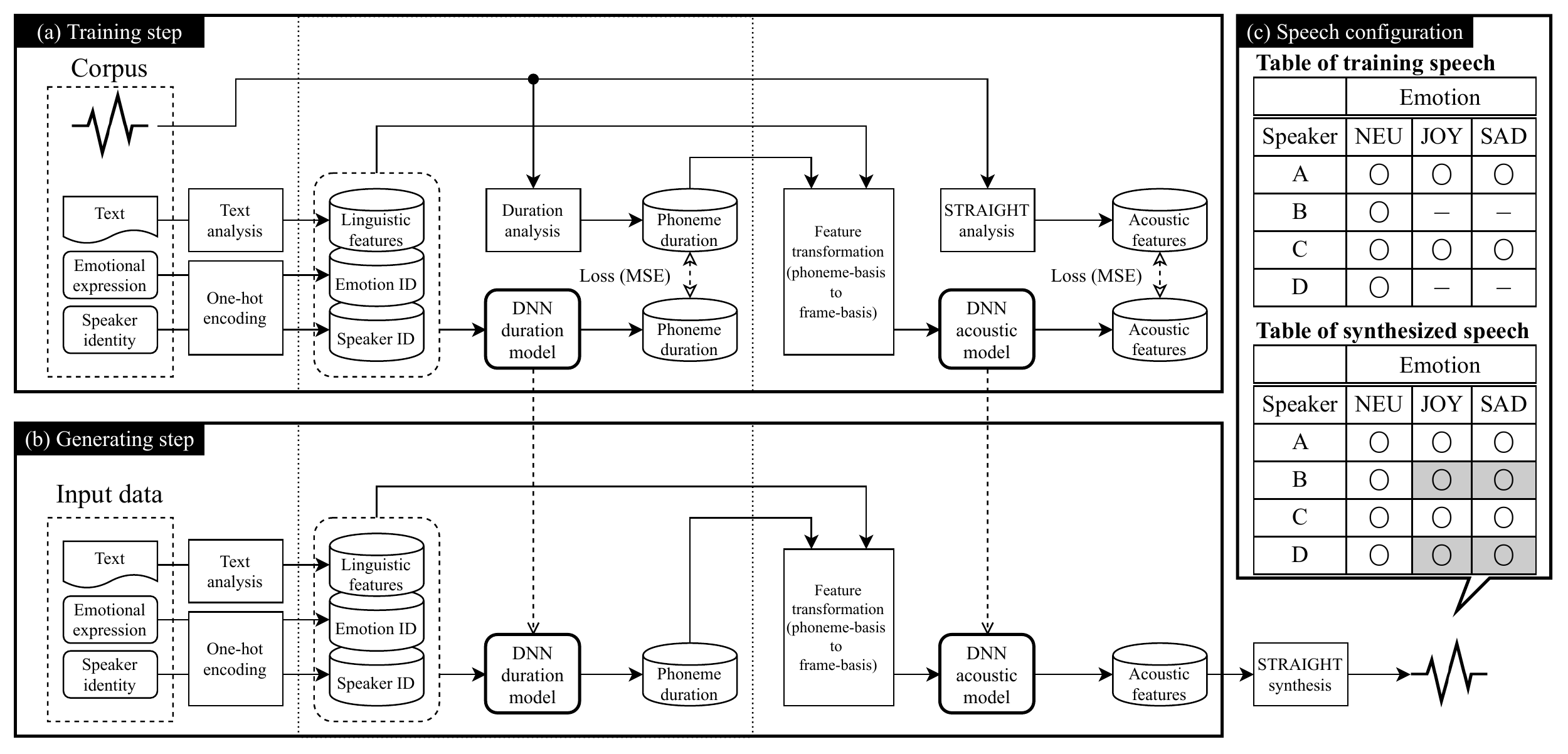}
			\caption{The proposed method of emotional speech synthesis in DNN-based TTS. In the table of training data, $\bigcirc$ indicates data used for training, --- indicates data not used for training. In the table of synthesized speech, $\bigcirc$ indicates data that the system can synthesize, hatching boxes indicate the extrapolation conditions, normal boxes indicate the interpolation conditions.}
			\label{fig:ProposedMethod}
			\end{center}
			\end{figure*}

\subsection{Expansion of output}

The outputs of $\ell$-th layer in eq.\ref{eq:DNN1} $\Vector{h}^{(\ell)}$ can be expanded as follows,
		\begin{align}
		\label{eq:DNN-OUT1}
		\Vector{h}^{(\ell)} 
		&= \left[  \Vector{h}^{(\ell)}_{(1)}~\Vector{h}^{(\ell)}_{(2)} \dots \Vector{h}^{(\ell)}_{(I)} 	\right]	\Vector{v}^{(\ell)}_{a} \\
		\label{eq:DNN-OUT2}
		\Vector{h}^{(\ell)}_{(i)} &= f^{(\ell)}_{(i)}( \Matrix{W}^{(\ell)}_{(i)} \Vector{h}^{(\ell-1)} + \Vector{b}^{(\ell)}_{(i)} )
		\end{align}
The output $\Vector{h}^{(\ell)}_{(i)} \in \Set{R}^{d_{\ell} \times 1}$ that corresponds to the $i$-th factor of auxiliary vector $v^{(\ell)}_{a(i)} \in \Set{R}$ is calculated from the shared input $\Vector{h}^{(\ell-1)}$.
The output of the $\ell$-th layer $\Vector{h}^{(\ell)}$ is a weighted sum of $\Vector{h}^{(\ell)}_{(i)}$ that uses the $i$-th factor of an auxiliary vector $v^{(\ell)}_{a(i)}$ as the weight.
When the auxiliary vector $\Vector{v}^{(\ell)}_{a}$ is a one-hot vector, the formula is related to DNN by using multitask learning \cite{bib-MTL}, which is a technique wherein a primary learning task is solved jointly with additional related tasks.
In multitask learning DNN \cite{bib-DNN-TTS-03}, the model has a shared hidden layer $\Vector{h}^{(\ell-1)}$ that can be considered the task-independent transformation.
Additionally, the model has multiple output layers corresponding to each task.

In multi-speaker DNN \cite{bib-DNN-MultiSpeaker-01} and emotional speech synthesis by speaker adaptation \cite{bib-DNN-MultiEmotion-03}, the model of each speaker has its own output layer, that is, the first task is speaker A, the second task is speaker B.

\section{Proposed method}\label{sec:Proposed}


\subsection{Overview of the proposed method}

Figure \ref{fig:ProposedMethod} presents the proposed DNN-based TTS that generates emotional speech by combining the emotional factor and the speaker factor.
In the training step presented in Fig. \ref{fig:ProposedMethod} (a), multi-speaker and multi-emotional speech data are used, where speakers and types of emotions are unbalanced.
That is, many speakers utter only neutral speech and few speakers utter both neutral and emotional speech. 
To synthesize emotional speech with the voice quality of the speakers who only utter neutral speech, DNNs must have an architecture to separately train the emotional factor and the speaker factor by introducing the auxiliary vectors. 
Details of the architecture are explained in \ref{subsec:Proposed_model}. 
In Fig. \ref{fig:ProposedMethod} (a), a phoneme duration model and acoustic model are trained. 
Both models have the same DNN architecture but different model parameters.
In the speech synthesis step presented in Fig. \ref{fig:ProposedMethod} (b), DNNs generate the target phoneme duration and the target acoustic features by setting speaker ID and emotion ID by using the auxiliary vectors.
In Fig. \ref{fig:ProposedMethod} (c), because any combination of speaker ID and emotion ID is possible, we synthesize emotional speech with the voice quality of the speakers who only utter neutral speech.

\subsection{Vector representation for controlling emotional expression and speaker identity}

Two types of a vector called emotion ID and speaker ID are used as features to control the emotional expression and speaker identity.
Several methods have been proposed to control speakers or emotions, for example, one-hot vector \cite{bib-DNN-MultiSpeaker-03_IEICE,bib-DNN-MultiEmotion-01}, i-vector \cite{bib-DNN-MultiSpeaker-02}, d-vector \cite{bib-d-vector}, x-vector \cite{bib-x-vector}.
Because the one-hot vector is simple and intuitive, we adopt its features to control emotions and speakers.

The emotion ID $E^{(i)}$ for the $i$-th emotion is defined as $E^{(i)} = \left[ e_1^{(i)}, e_2^{(i)}, ..., e_M^{(i)} \right]^\top$, where each value $e_{m}^{(i)}$ is expressed as follows:
	\begin{equation}
		e_{m}^{(i)} = \Vector{1}_{m=i}
	\label{eq:EMcode}
	\end{equation}
	%
where $M$ is the dimension of $E^{(i)}$ and equal to the number of emotions in the training data.
Additionally, $\Vector{1}_{m=i}$ is $1$ if $m=i$ is true, and $0$ otherwise.
To represent the neutral emotion, the emotion ID is a zero vector. 

In the same manner, the speaker ID $S^{(j)}$ for the $j$-th speaker is defined as $S^{(j)} = \left[ s_1^{(j)}, s_2^{(j)}, ..., s_N^{(j)} \right]^\top$, where each value $s_{n}^{(j)}$ is expressed as follows:
	\begin{equation}
		s_{n}^{(j)} = \Vector{1}_{n=j}
	\label{eq:SPcode}
	\end{equation}
	%
where $N$ is the dimension of $S^{(j)}$ and equal to the number of speakers in the training data. 

		\begin{figure}[bt]
			\begin{center}
			\includegraphics[width=\columnwidth]{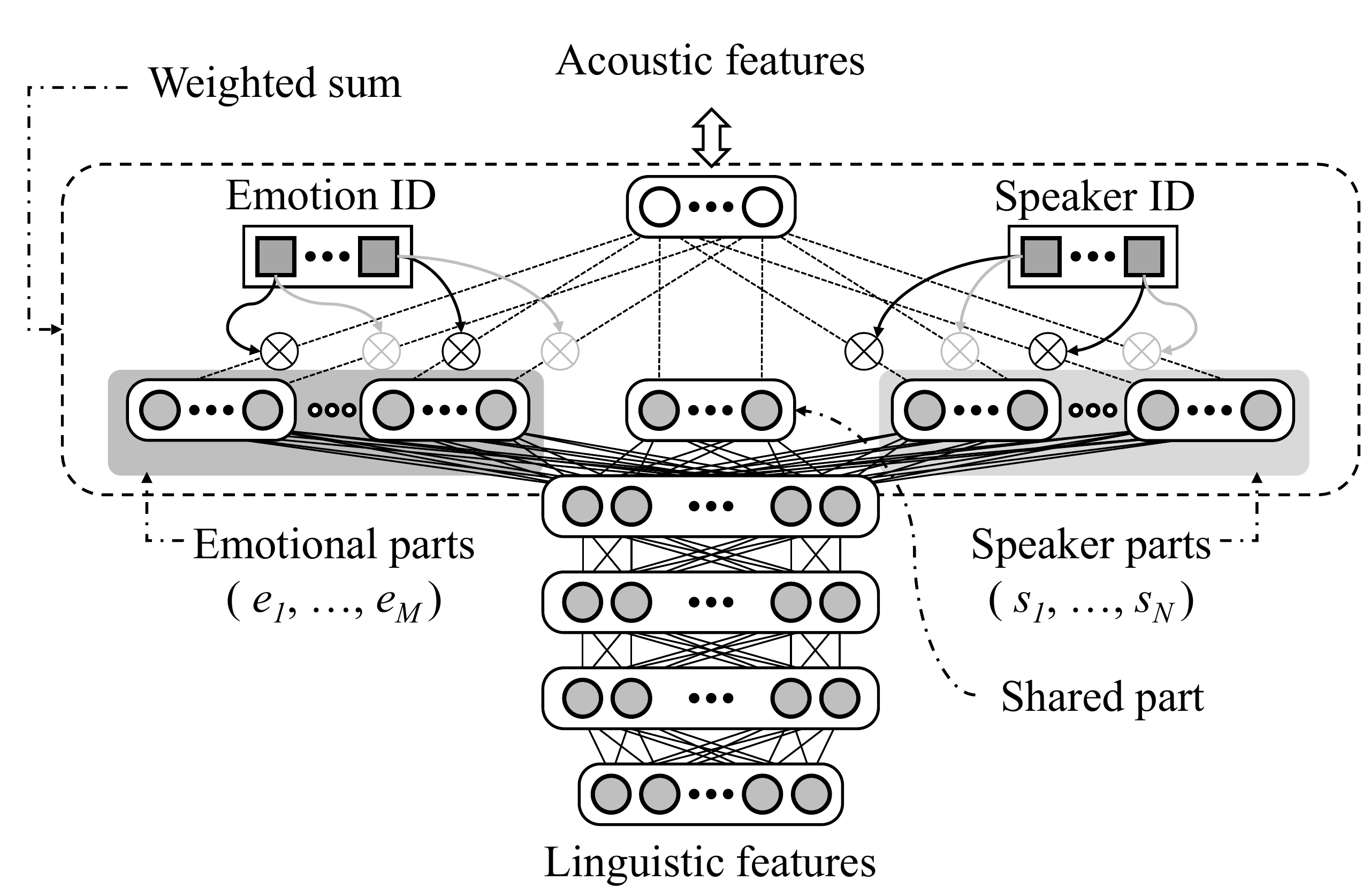}
			\caption{The parallel model (PM)}
			\label{fig:PM}
			\end{center}
		\end{figure}

	\begin{table*}[tb]
\caption{Data configuration for training and evaluation. (In the model training, $\bigcirc$ indicates that data was included, and --- indicates that data was not included. In the evaluation, the hatching box indicates the synthesized speech used for the evaluation. N indicates neutral, J indicates joyful, and S indicates sad.)}
	\label{tab:data_conf}
	\vspace{5pt}
	\centering
	\begin{tabular}{l|ccc|ccc|ccc|ccc|ccc}
	\hline
	\hline
	\multirow{3}{*}{\begin{tabular}{c}Type of synthesized speech \\for the evaluation experiment \end{tabular}}
	&\multicolumn{12}{c|}{corpus $\alpha$}
	&\multicolumn{3}{c}{corpus $\beta$} \\	
	\cline{2-16}
	&\multicolumn{3}{c|}{Female A}
	&\multicolumn{3}{c|}{Female B}
	&\multicolumn{3}{c|}{Male A}
	&\multicolumn{3}{c|}{Male B}
	&\multicolumn{3}{c}{12 speakers} \\
	\cline{2-16}
	&N						&J 											&S
	&N						&J											&S
	&N						&J 											&S
	&N						&J 											&S
	&N						&J 											&S \\
	\hline
	(a)\begin{tabular}{l}Open-emotion test of female A \end{tabular}
	&\small{$\bigcirc$}	&\cellcolor[gray]{0.8}\small{---}			&\cellcolor[gray]{0.8}\small{---}
	&\small{$\bigcirc$}	&\small{$\bigcirc$}						&\small{$\bigcirc$}
	&\small{$\bigcirc$}	&\small{$\bigcirc$}						&\small{$\bigcirc$}
	&\small{$\bigcirc$}	&\small{---} 								&\small{---}
	&\small{$\bigcirc$}	&\small{---}								&\small{---} \\
	(b)\begin{tabular}{l}Open-emotion test of female B \end{tabular}
	&\small{$\bigcirc$}	&\small{$\bigcirc$}						&\small{$\bigcirc$}
	&\small{$\bigcirc$}	&\cellcolor[gray]{0.8}\small{---}			&\cellcolor[gray]{0.8}\small{---}
	&\small{$\bigcirc$}	&\small{$\bigcirc$}						&\small{$\bigcirc$}
	&\small{$\bigcirc$}	&\small{---}								&\small{---}
	&\small{$\bigcirc$}	&\small{---}								&\small{---} \\
	\hline
	(c)\begin{tabular}{l}Closed-emotion test of female A/B  \end{tabular}
	&\small{$\bigcirc$}	&\cellcolor[gray]{0.8}\small{$\bigcirc$}	&\cellcolor[gray]{0.8}\small{$\bigcirc$}
	&\small{$\bigcirc$}	&\cellcolor[gray]{0.8}\small{$\bigcirc$}	&\cellcolor[gray]{0.8}\small{$\bigcirc$}
	&\small{$\bigcirc$}	&\small{$\bigcirc$}						&\small{$\bigcirc$}
	&\small{$\bigcirc$}	&\small{---}								&\small{---}
	&\small{$\bigcirc$}	&\small{---}								&\small{---} \\
	\hline
	(d)\begin{tabular}{l}Speaker-and-emotion-dependent (SED)\\test of female A's joyful \end{tabular}
	&\small{---}			&\cellcolor[gray]{0.8}\small{$\bigcirc$} \cellcolor[gray]{0.8}	&\small{---}
	&\small{---}			&\small{---}								&\small{---}
	&\small{---}			&\small{---}								&\small{---}
	&\small{---}			&\small{---}								&\small{---}
	&\small{---}			&\small{---}								&\small{---} \\
	(e)\begin{tabular}{l}SED test of female A's sad	\end{tabular}
	&\small{---}			&\small{---} 								&\cellcolor[gray]{0.8}\small{$\bigcirc$}
	&\small{---}			&\small{---} 								&\small{---}
	&\small{---}			&\small{---}								&\small{---}
	&\small{---}			&\small{---}								&\small{---}
	&\small{---}			&\small{---}								&\small{---} \\
	(f)\begin{tabular}{l}SED test of female B's joyful \end{tabular}
	&\small{---}			&\small{---} 								&\small{---}
	&\small{---}			&\cellcolor[gray]{0.8}\small{$\bigcirc$}	&\small{---}
	&\small{---}			&\small{---} 								&\small{---}
	&\small{---}			&\small{---}								&\small{---}
	&\small{---}			&\small{---}								&\small{---} \\
	(g)\begin{tabular}{l}SED test of female B's sad	\end{tabular}
	&\small{---}			&\small{---} 								&\small{---}
	&\small{---}			&\small{---}								&\cellcolor[gray]{0.8}\small{$\bigcirc$}
	&\small{---}			&\small{---}								&\small{---}
	&\small{---}			&\small{---}								&\small{---}
	&\small{---}			&\small{---}								&\small{---} \\
	\hline
	\hline
	\end{tabular}
	\end{table*}

\subsection{Proposed model architecture}
\label{subsec:Proposed_model}

We propose five types of DNNs that can separately control the speaker factor and emotional factor.

\subsubsection{Parallel model}

In Fig. \ref{fig:PM}, a PM has an output layer comprised of emotion-dependent parts (Emotion 1, Emotion 2, ..., Emotion M), speaker-dependent parts (Speaker 1, Speaker 2, ..., Speaker N), and a shared part.
The PM uses $\left[ E^{(i)\top} \: S^{(j)\top} \: 1 \right]^\top$ as the auxiliary vector $\Vector{v}^{(\ell)}_{a}$ in the output layer $\ell=L$ at eq.\ref{eq:DNN-OUT1}.
The outputs of the emotion-dependent part and speaker-dependent part are summed linearly, because the linear activation function is used at the output layer.
The PM is newly proposed and is motivated by a multi-speaker DNN \cite{bib-DNN-MultiSpeaker-01} and the emotion additive model \cite{bib-HMM-MultiEmotion-open-03}, where hidden layers are regarded as a linguistic feature transformation shared by all speakers \cite{yamagishi2003average_voice}.
Because the acoustic feature is represented as the addition of the emotional-dependent part, and the speaker-dependent part, the emotional factor and speaker factor are separately controlled.

\subsubsection{Serial model}

In a SM, the speaker factor and emotional factor are sequentially modeled in different layers.
The two types of architectures are: that the SM$_{\mathrm{se}}$, which models the speaker factor in the former layer, and then models the emotional factor in the later layer: the other type of architecture is the SM$_{\mathrm{es}}$, which models the two factors in reverse order.
The SM$_{\mathrm{se}}$ uses $\left[ S^{(j)\top} \: 1 \right]^\top$ as the auxiliary vector $\Vector{v}^{(\ell)}_{a}$ in the last hidden layer $\ell=L-1$ at eq.\ref{eq:DNN-OUT1}, and $\left[ E^{(i)\top} \: 1 \right]^\top$ as the auxiliary vector $\Vector{v}^{(\ell)}_{a}$ in the output layer $\ell=L$ at eq.\ref{eq:DNN-OUT1}.
Opposite to SM$_{\mathrm{se}}$, SM$_{\mathrm{es}}$ uses $\left[ E^{(i)\top} \: 1 \right]^\top$ in the last hidden layer, and $\left[ S^{(j)\top} \: 1 \right]^\top$ in the output layer.
As with the PM, in the output layer, the output of the speaker-dependent part or emotion-dependent part is summed linearly.
However, in the hidden layer, the output of the emotion-dependent part or speaker-dependent part is summed nonlinearly because the sigmoid activation function is used.

\subsubsection{Auxiliary input model}

An AIM implicitly models the speaker factor and emotional factor by forwarding the value of the auxiliary input vector.
The AIM uses $\left[ E^{(i)\top} \: S^{(j)\top} \right]^\top$ in the auxiliary vector $\Vector{v}^{(\ell)}_{a}$ of the input layer $\ell=0$ in eq.\ref{eq:DNN-IN2}.

\subsubsection{Hybrid models}

As hybrid models, two types of model are used, namely the combination of PM and AIM (PM\&AIM), and the combination of SM and AIM (SM\&AIM).
The PM\&AIM uses the auxiliary vector if the input layer $\ell=0$ and the output layer $\ell=L$.
Also, the SM\&AIM uses the auxiliary vector if the input layer $\ell=0$, the last hidden layer $\ell=L-1$ and the output layer $\ell=L$.

\section{Evaluation experiments}\label{sec:Experiments}

To evaluate the extrapolation performance of the proposed architectures, open-emotion tests and closed-emotion tests are conducted objectively and subjectively. 
Here, open-emotion and closed-emotion mean that, in the training step, the emotional speech of a target speaker is not included and is included, respectively.

	\begin{table*}[tb]
	\caption{The DNN architectures for the phoneme duration model and the acoustic model. (The hatching box indicates the expanded part of the simplest DNN.)}
	\label{tab:model}
	\vspace{5pt}
	\centering
	\resizebox{\textwidth}{!}{
	
	\begin{tabular}{l|cccc|ccccc}
	\hline
	&\multicolumn{4}{c|}{(a) Phoneme duration model}	&\multicolumn{5}{c}{(b) Acoustic model}\\
	\cline{2-10}
	Model
	&\raisebox{-10pt}{{\footnotesize \shortstack{Input \\ layer}}}
	&\raisebox{-10pt}{{\footnotesize \shortstack{Hidden \\ layer 1}}}
	&\raisebox{-10pt}{{\footnotesize \shortstack{Hidden \\ layer 2}}}
	&\raisebox{-10pt}{{\footnotesize \shortstack{Output \\ layer}}}
	&\raisebox{-10pt}{{\footnotesize \shortstack{Input \\ layer}}}
	&\raisebox{-10pt}{{\footnotesize \shortstack{Hidden \\ layer 1}}}
	&\raisebox{-10pt}{{\footnotesize \shortstack{Hidden \\ layer 2}}}
	&\raisebox{-10pt}{{\footnotesize \shortstack{Hidden \\ layer 3}}}
	&\raisebox{-10pt}{{\footnotesize \shortstack{Output \\ layer}}}\\
	\hline
	PM
	&298	&32	&32&\cellcolor[gray]{0.8}\textbf{1*(16+2+1)}
	&305	&256	&256	&256	&\cellcolor[gray]{0.8}\textbf{154*(16+2+1)}	\\
	SM$_{\mathrm{se}}$
	&298	&32	&\cellcolor[gray]{0.8}\textbf{32*(16+1)}	&\cellcolor[gray]{0.8}\textbf{1*(2+1)}
	&305	&256	&256	&\cellcolor[gray]{0.8}\textbf{256*(16+1)}	&\cellcolor[gray]{0.8}\textbf{154*(2+1)} \\
	SM$_{\mathrm{es}}$
	&298	&32	&\cellcolor[gray]{0.8}\textbf{32*(2+1)}	&\cellcolor[gray]{0.8}\textbf{1*(16+1)}
	&305	&256	&256	&\cellcolor[gray]{0.8}\textbf{256*(2+1)}	&\cellcolor[gray]{0.8}\textbf{154*(16+1)} \\
	AIM
	&\cellcolor[gray]{0.8}\textbf{298+16+2}	&64	&64	&1
	&\cellcolor[gray]{0.8}\textbf{305+16+2}	&512	&512	&512	&154 \\
	PM\&AIM
	&\cellcolor[gray]{0.8}\textbf{298+16+2}	&32	&32&\cellcolor[gray]{0.8}\textbf{1*(16+2+1)}
	&\cellcolor[gray]{0.8}\textbf{305+16+2}	&256	&256	&256	&\cellcolor[gray]{0.8}\textbf{154*(16+2+1)}	\\
	SM$_{\mathrm{se}}$\&AIM
	&\cellcolor[gray]{0.8}\textbf{298+16+2}	&32	&\cellcolor[gray]{0.8}\textbf{32*(16+1)}	&\cellcolor[gray]{0.8}\textbf{1*(2+1)}
	&\cellcolor[gray]{0.8}\textbf{305+16+2}	&256	&256	&\cellcolor[gray]{0.8}\textbf{256*(16+1)}	&\cellcolor[gray]{0.8}\textbf{154*(2+1)} \\
	SM$_{\mathrm{es}}$\&AIM
	&\cellcolor[gray]{0.8}\textbf{298+16+2}	&32	&\cellcolor[gray]{0.8}\textbf{32*(2+1)}	&\cellcolor[gray]{0.8}\textbf{1*(16+1)}
	&\cellcolor[gray]{0.8}\textbf{305+16+2}	&256	&256	&\cellcolor[gray]{0.8}\textbf{256*(2+1)}	&\cellcolor[gray]{0.8}\textbf{154*(16+1)} \\
	SED
	&298	&32	&32	&1
	&305	&256	&256	&256	&154 \\
	\hline
	\end{tabular}
	}
	\end{table*}

 \begin{table}[tb]
	\caption{Confusion matrices for subjective emotional classification results (Value indicates the accuracy of classification. * indicates p$<0.001$ in a chi-square test between the closed-emotion by the PM (f) and the others (a, b, c, d, and e).)}
	\label{tab:ex4_acc}
	\vspace{5pt}
	\centering
	\begin{tabular}{l|llll}
	\multicolumn{5}{c}{(a) Open-emotion by the PM}\\
	\hline
	Correct	&\multicolumn{4}{c}{Judged emotion} \\
	\cline{2-5}
	emotion	&NEU				&JOY						&SAD			&OTH \\
	\hline
	NEU		&\textbf{0.87}		&0.08						&0.05			&0.01 \\
	JOY		&0.31				&\textbf{0.61}				&0.03			&0.05 \\
	SAD		&0.32				&0.02						&\textbf{0.65}	&0.02 \\
	\hline
	\multicolumn{5}{c}{}\\
	\multicolumn{5}{c}{(b) Open-emotion by the PM\&AIM}\\
	\hline
	Correct	&\multicolumn{4}{c}{Judged emotion} \\
	\cline{2-5}
	emotion	&NEU				&JOY						&SAD			&OTH \\
	\hline
	NEU		&\textbf{0.85}		&0.11						&0.04			&0.00 \\
	JOY		&0.38				&\textbf{0.54}$^{*}$		&0.05			&0.03 \\
	SAD		&0.30				&0.02						&\textbf{0.68}	&0.01 \\
	\hline
	\multicolumn{5}{c}{}\\
	\multicolumn{5}{c}{(c) Open-emotion by the AIM}\\
	\hline
	Correct	&\multicolumn{4}{c}{Judged emotion} \\
	\cline{2-5}
	emotion	&NEU				&JOY						&SAD			&OTH \\
	\hline
	NEU		&\textbf{0.89}		&0.06						&0.04			&0.01 \\
	JOY		&0.44				&\textbf{0.42}$^{*}$		&0.08			&0.06 \\
	SAD		&0.46				&0.01						&\textbf{0.53}	&0.01 \\
	\hline
	\multicolumn{5}{c}{}\\
	\multicolumn{5}{c}{(d) Open-emotion by the SM$_{\mathrm{es}}$}\\
	\hline
	Correct	&\multicolumn{4}{c}{Judged emotion} \\
	\cline{2-5}
	emotion	&NEU				&JOY						&SAD			&OTH \\
	\hline
	NEU		&\textbf{0.87}		&0.09						&0.03			&0.01 \\
	JOY		&0.65				&\textbf{0.24}$^{*}$		&0.09			&0.02 \\
	SAD		&0.29				&0.02						&\textbf{0.66}	&0.03 \\
	\hline
	\multicolumn{5}{c}{}\\
	\multicolumn{5}{c}{(e) Open-emotion by the SM$_{\mathrm{es}}$\&AIM}\\
	\hline
	Correct	&\multicolumn{4}{c}{Judged emotion} \\
	\cline{2-5}
	emotion	&NEU				&JOY						&SAD			&OTH \\
	\hline
	NEU		&\textbf{0.87}		&0.10						&0.03			&0.00 \\
	JOY		&0.55				&\textbf{0.40}$^{*}$		&0.04			&0.01 \\
	SAD		&0.29				&0.02						&\textbf{0.68}	&0.01 \\
	\hline
	\multicolumn{5}{c}{}\\
	\multicolumn{5}{c}{(f) Closed-emotion by the PM}\\
	\hline
	Correct	&\multicolumn{4}{c}{Judged emotion} \\
	\cline{2-5}
	emotion	&NEU				&JOY						&SAD			&OTH \\
	\hline
	NEU		&\textbf{0.89}		&0.08						&0.02			&0.01 \\
	JOY		&0.22				&\textbf{0.73}				&0.02			&0.03 \\
	SAD		&0.44				&0.03						&\textbf{0.51}	&0.02 \\
	\hline
	\end{tabular}
	\end{table}

\subsection{Emotional speech database}
\label{sec:DB}
\subsubsection{Overview of speech database}

In the experiments, two types of Japanese speech corpus, corpus $\alpha$ and $\beta$, were used.
In corpus $\alpha$, the same $500$ sentences were uttered in several ways.
Two female speakers and a male speaker uttered the sentences with neutral, joyful, and sad emotions; and a male speaker uttered them with only neutral emotion.
Each dataset's duration of each dataset was approximately $35$ minutes, and the duration of the entire ``corpus $\alpha$'' is approximately $350$ minutes.
In corpus $\beta$, the same $130$ sentences that differed from corpus $\alpha$, were uttered with only neutral emotion by six female and six male speakers.
The duration of each dataset was approximately $40$ minutes, and the duration of the entire corpus $\beta$ is approximately $480$ minutes.
The speech signals were sampled at $22.05$ kHz and quantized at $16$ bits.
The phoneme duration was manually annotated and labeled with the same format as the hidden Markov model toolkit (HTK)~\cite{young2006htk}.


Speech data were divided into a training-set, validation-set, and test-set with a rate of 90\%:5\%:5\%, that is, corpus $\alpha$ was divided into 450:25:25, and corpus $\beta$ was divided into 120:5:5.

\subsubsection{Training data combination and test data}


Table \ref{tab:data_conf} presents a summary of training and test data. 
A circle in the table indicates that $95$\% of the data (the training-set) is used for training, and a dash in the table indicates that speech is not included in the training. 
Hatching in the table indicates data used for the evaluation. 
A circle in the hatching box indicates that $5$\% of the data (the test-set) is used for evaluation. 
For the open-emotion test, DNN was trained using data (a) and (b) in Table \ref{tab:data_conf}. 
The synthesized speech is compared with the real emotional speech uttered by female speakers A and B. 
For the closed-emotion test, DNN is trained using data (c) in Table \ref{tab:data_conf}. 
The synthesized speech is compared with the real emotional speech uttered by female speakers A and B. 
As references, speaker-and-emotion-dependent models (SEDs) were trained using the training-set of (d), (e), (f), and (g), and evaluated using their test-sets.

%


\subsection{Model training}

The proposed models and SEDs are trained using the database described in \ref{sec:DB}. 
The STRAIGHT \cite{bib-STRAIGHT} analysis is used to extract the spectral envelop, aperiodicity, F0, and voiced/unvoiced flag in a $5$-ms frame shift.
Next, $40$-dimensional Mel-cepstral coefficients, $10$-band-aperiodicities, and F0 in log-scale are calculated. 
Notably, $80$\% of the silent frames are removed from the training data to avoid increasing the proportion of silence in the training data and reduce the computational cost.



\subsubsection{Phoneme duration model}
\label{subsubsec:duration}

For the PM, SM, and SED, the input feature vectors are $289$-dimensional binary features of categorical linguistic contexts (e.g., quinphone, the interrogative sentence flag), and $9$-dimensional numerical linguistic contexts (e.g., the number of mora in the current word, the relative position toward the accent nucleus in the current mora).
For the AIM, PM\&AIM, and SM\&AIM, as auxiliary features, speaker and emotion IDs are added to the input feature vectors.
Because the speaker ID is $16$ dimensions and the emotion ID is $2$ dimensions, the dimension of the input vector becomes $316$.

The output feature is the integer scalar value that indicates the number of frames (i.e., phoneme duration).
The output features of the training data are normalized to zero mean and unit variance.

As DNN model architectures, the FFNNs presented in Table \ref{tab:model} (a) are used.
A sigmoid function is used in the hidden layers followed by a linear activation at the output layer.
For the training process, the weights of all DNN (PM, SM, AIM, PM\&AIM, SM\&AIM and SED) are randomly initialized.
The weights are trained using a backpropagation procedure with a minibatch-based MomentumSGD to minimize the mean squared error between the output features of the training data and the predicted values.
The initial learning rate of MomentumSGD is $0.16$ (PM, SM, PM\&AIM, SM\&AIM, and SED) or $0.08$ (AIM), and the momentum is $0.9$.
The training data for the minibatch is randomly selected, and the minibatch size is $64$ (PM, SM, AIM, PM\&AIM, and SM\&AIM) or $16$ (SED).
The schedule of the training is a similar method, to randomly select the data as conventional DNN.
The hyper-parameters of each model were selected by a grid search that has a higher performance based on a smaller number of parameters.

\subsubsection{Acoustic model}

For the PM, SM, and SED, $7$-dimensional time features (e.g., the total frame number of current mora, the current state in $5$-state) are added to the feature vector used in the phoneme duration model.
Thus, the dimension of the input feature vector is $305$.
The dimension of the input feature vectors for the AIM, PM\&AIM, and SM\&AIM, is $323$ because both the speaker and emotion IDs are added as auxiliary features in the same manner as in the phoneme duration model.

The output feature vector contains log F0, $40$ Mel-cepstral coefficients, $10$-band-aperiodicities, their delta and delta-delta counterparts, and a voiced/unvoiced flag, which results in $154$ dimensions.
The voiced/unvoiced flag is a binary feature that indicates the voicing of the current frame.
The output features of the training data are normalized to zero mean and unit variance.
In these experiments, phoneme durations extracted from natural speech are used.

As DNN model architectures, the FFNNs presented in Table \ref{tab:model} (b) were used.
Activation function, loss function, optimizer, and training schedules are the same condition as in Section \ref{subsubsec:duration}.
The initial learning rate of MomentumSGD is $1.28$ (PM, SM, AIM, PM\&AIM, SM\&AIM, and SED), and the momentum is $0.9$.
The training data for the minibatch is randomly selected, and the minibatch size is $128$ (PM, SM, AIM, PM\&AIM, SM\&AIM and SED). 
The hyper-parameters of each model are selected by using the same rule we used with the phoneme duration model.

\subsection{Objective evaluation experiments}

To make sure the advantages of the proposed architectures, objective evaluations were performed by comparing the estimated values using the models with extracted values from real emotional speech. 
Another aim of the experiments is to know the upper limit of interpolation by comparing the estimated values of the open-emotion test with those of the closed-emotion test.


\begin{figure}[t]
\begin{center}
\includegraphics[width=0.9\columnwidth]{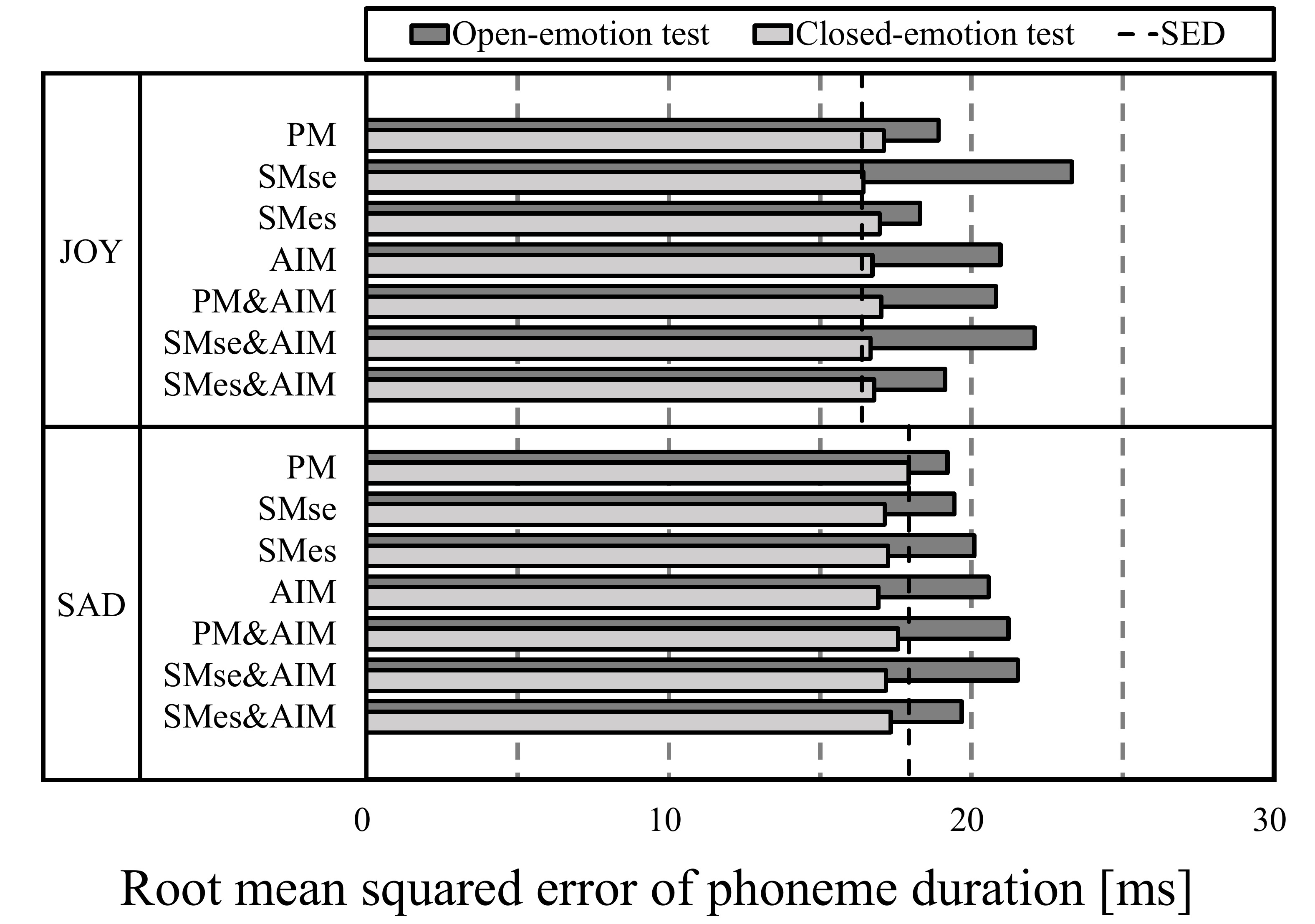}
\vspace{-15pt}     
\end{center}
\caption{Objective evaluation results of the RMSE of phoneme duration}
\label{fig:dur_dist}
\end{figure}
\begin{figure}[t]
	\begin{center}
		\includegraphics[width=0.9\columnwidth]{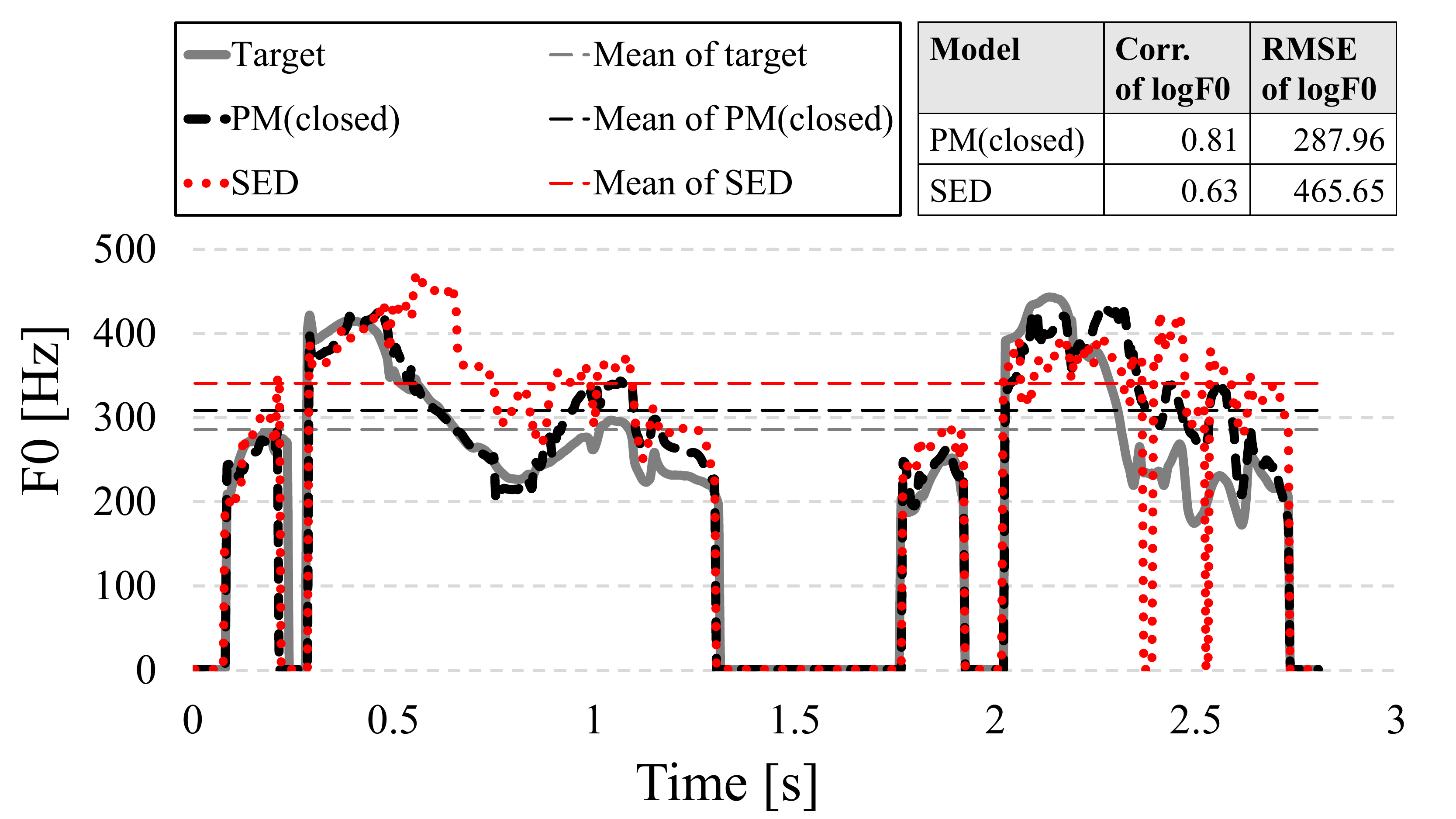}
		\vspace{-15pt}     
		\end{center}
		\caption{The comparison of F0 of female B's joyful speech. The upper right table shows the values that are calculated using the shown F0 patterns.}
		\label{fig:logF0_compare}
	\end{figure}

\subsubsection{Objective evaluation of the phoneme duration model}

The phoneme duration models are evaluated by the root mean squared error (RMSE) of phoneme duration that is calculated between the phoneme duration extracted from real emotional speech and the phoneme duration generated by the phoneme duration model.

Figure \ref{fig:dur_dist} presents the experimental results.
Compared with the open-emotion test and the closed-emotion test, the difference is less than $5$ ms except for the joyful speech of SM$_{\mathrm{se}}$ and SM$_{\mathrm{se}}$\&AIM. 
The differences are small and might not be perceivable by hearing. 
Based on the results, for phoneme duration modeling, the proposed model works well, and we do not collect emotional speech uttered by each speaker.


\subsubsection{Objective evaluation of the acoustic model}

The acoustic models are evaluated by the correlation coefficient of log F0, the RMSE of log F0, and the Mel-cepstral distortion (MCD).
The objective measure is calculated between the parameters extracted from the real emotional speech and the parameters generated by the acoustic model.
Figure \ref{fig:logF0_compare} shows the F0 contours and the average F0 of joyful speech that is generated by the SED and PM in the closed-emotion test as well as those extracted from the target speaker’s joyful speech. 
The upper right table indicates the correlation coefficient of log F0 and the RMSE of log F0 calculated from this single utterance.
It is observed that the correlation coefficient of log F0 and the RMSE of log F0 are useful for evaluating whether the F0 contour and average F0 are similar to the target speech.

\begin{figure}[t]
\begin{center}
\includegraphics[width=0.9\columnwidth]{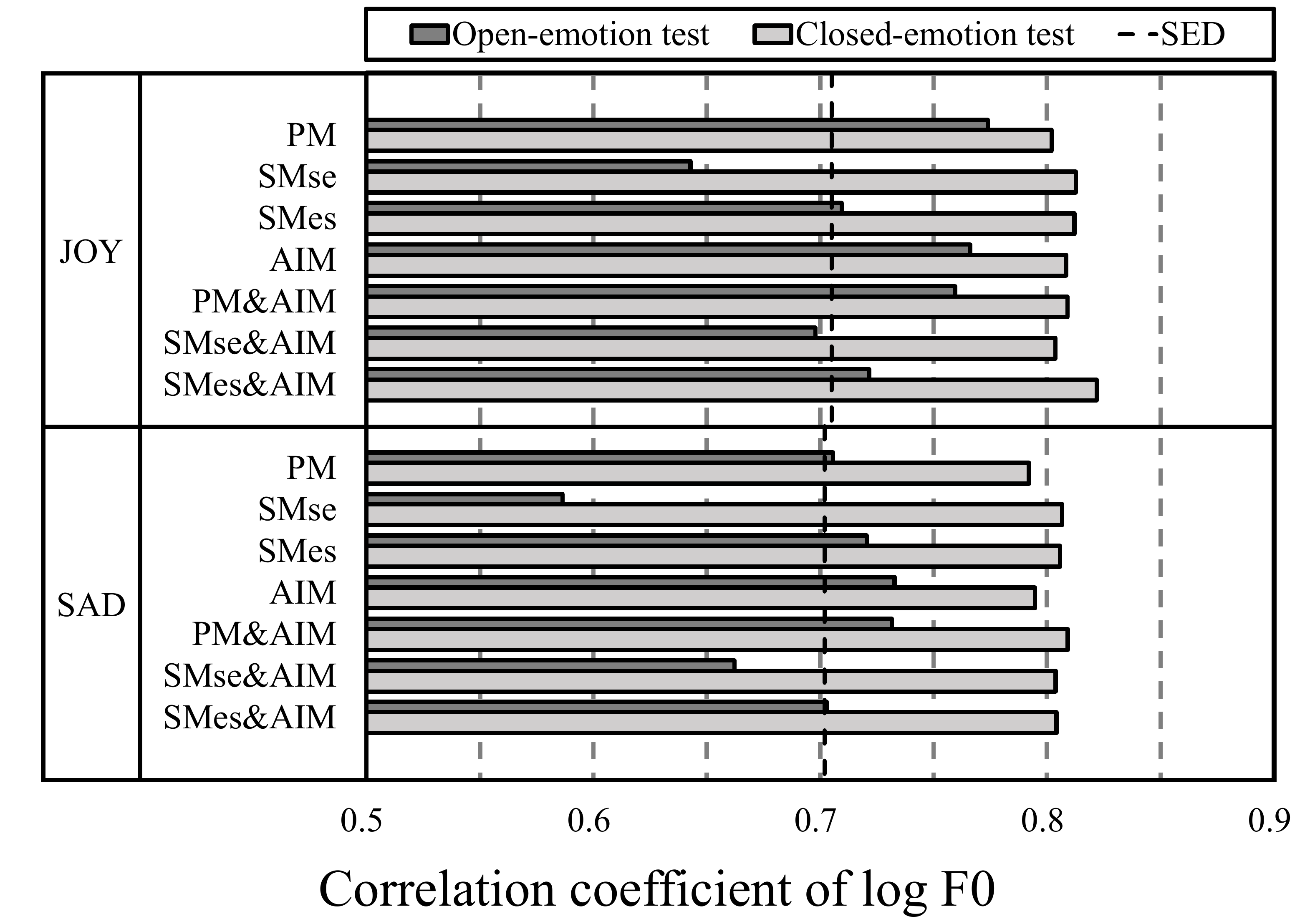}
\vspace{-15pt}     
\end{center}
\caption{Objective evaluation results of  the correlation coefficient of log F0}
\label{fig:logF0_corr}
\begin{center}
\includegraphics[width=0.9\columnwidth]{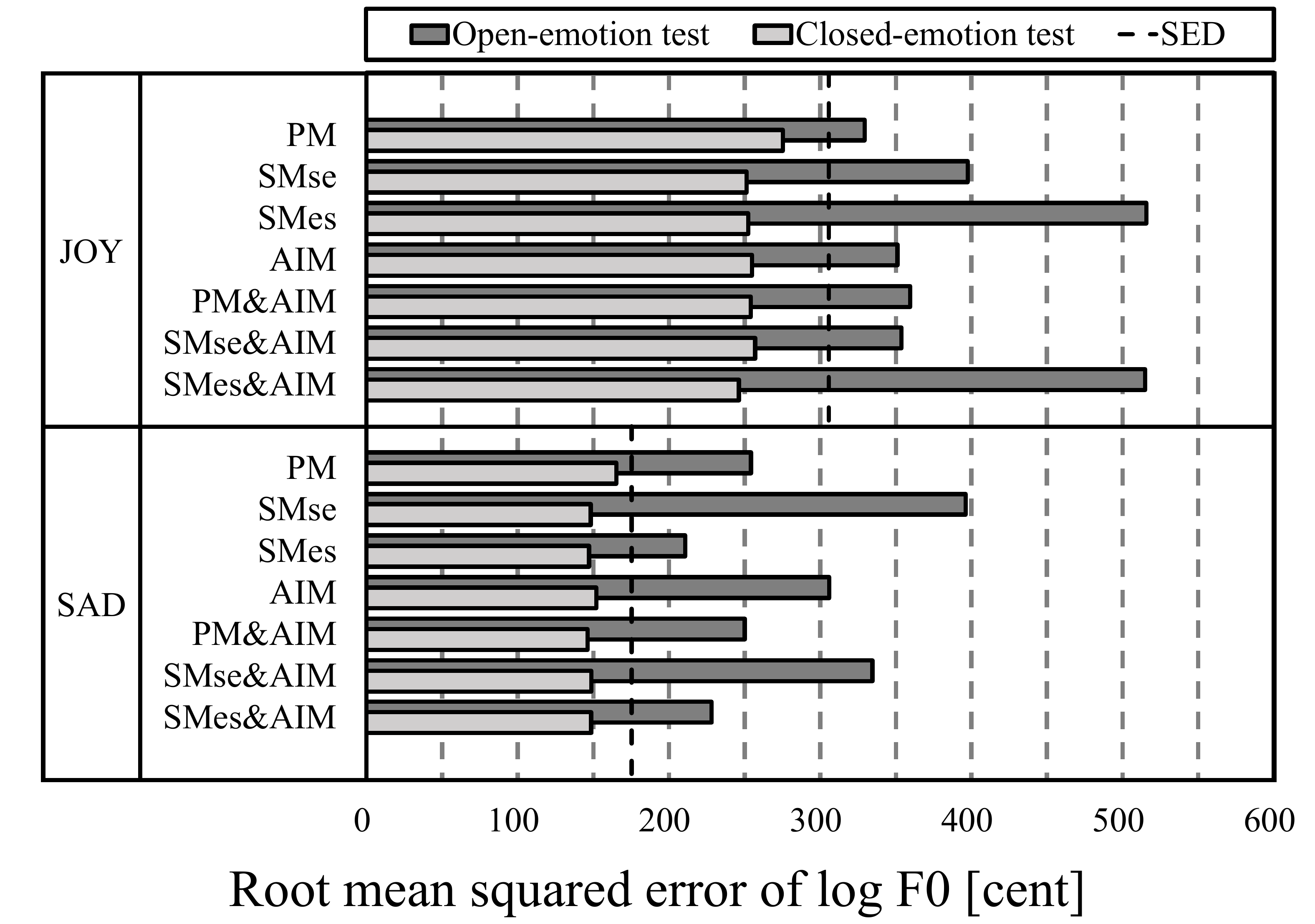}
\vspace{-15pt}     
\end{center}
\caption{Objective evaluation results of the RMSE of log F0}
\label{fig:logF0_dist}
\end{figure}

Figure \ref{fig:logF0_corr} presents the results for the correlation coefficient of log F0.
In closed-emotion test, the SED has poorer performance than the proposed models. 
The main reason for the results is that the SED takes approximately $35$ minutes of training data, while the proposed model takes approximately $760$ minutes. 
Moreover, corpus $\beta$ contains different texts from corpus $\alpha$, which results in increasing variations in phoneme contexts. 
This is another advantage of the proposed approach; i.e., we can effectively use speech data from many speakers. 
Interestingly, even in the open-emotion test, all the proposed models except the SM$_{\mathrm{se}}$ and SM$_{\mathrm{se}}$\&AIM have the same or better performance compared with the SED. 
This is because the all speakers uttered the same text with several emotions. 
Because Japanese is a tonal language, F0 patterns are important to convey meanings.
So speakers cannot drastically change the shape of F0 patterns, but can change only the height or length of F0 patterns to express emotions. 
Therefore, the correlation coefficient of log F0 showed fairly good in open-emotion tests. 
The degradation of the coefficient is only $0.1$ from the closed-emotion test.

Figure \ref{fig:logF0_dist} presents the results for the RMSE of log F0.
In contrast to the correlation coefficient of log F0, in the open-emotion test, all the proposed models have poorer performances than the SED. 
This mainly occurs because each speaker has their own means to control F0 contours to express emotion.

\begin{figure}[t]
\begin{center}
\includegraphics[width=0.9\columnwidth]{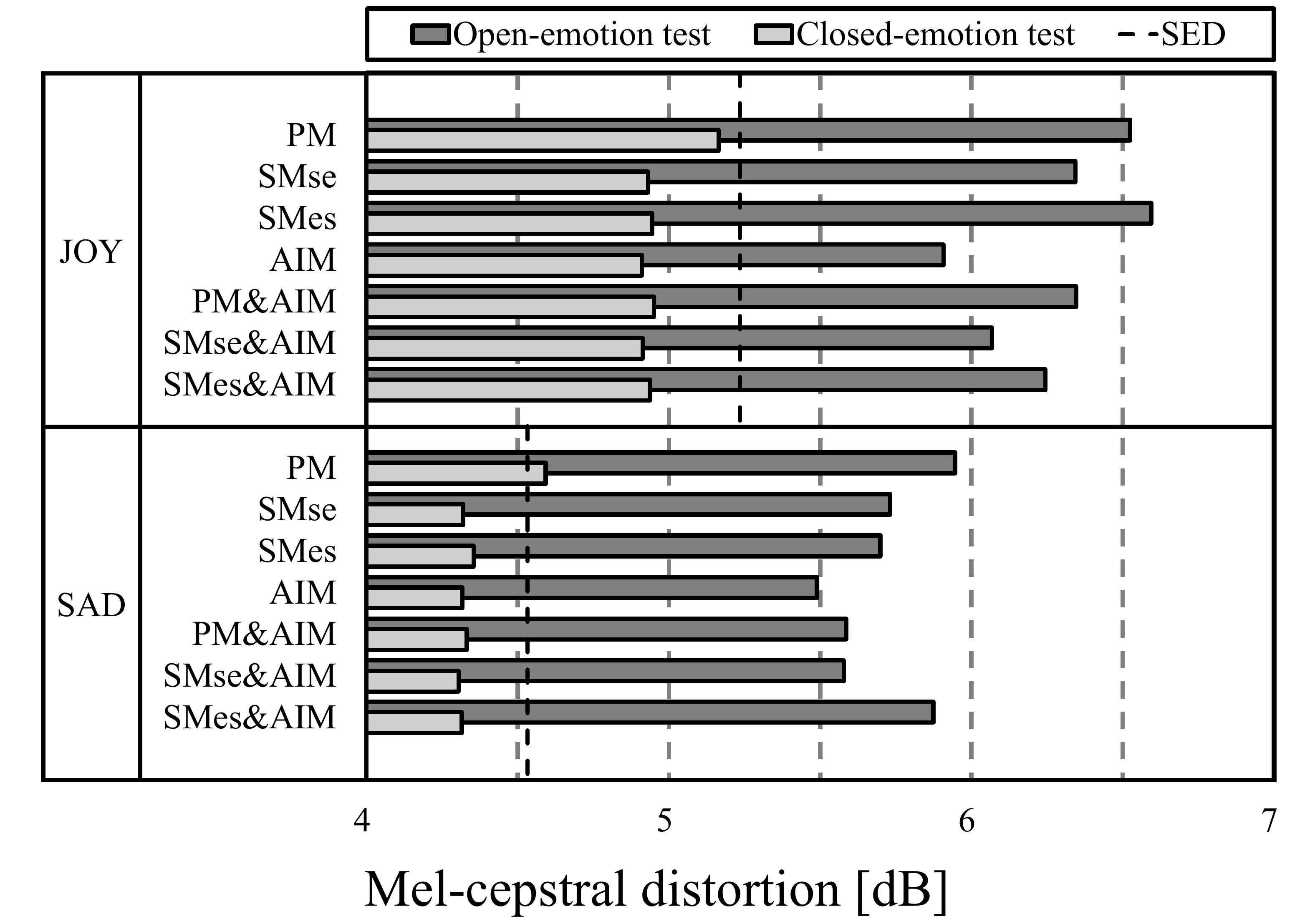}
\vspace{-15pt}     
\end{center}
\caption{Objective evaluation results of the MCD}
\label{fig:mcep_dist}
\end{figure}

Figure \ref{fig:mcep_dist} presents the results for the MCD.
In the open-emotion test, all the proposed models again have poorer performances than the SED. 
This indicates that each speaker also changes their articulation in their own fashion to express emotion and the F0 contour.

In terms of the overall performance of the models, we can say that SM$_{\mathrm{se}}$ and SM$_{\mathrm{es}}$ are not promising, because their performance showed different tendencies and is not predictable.
This was shown in the correlation coefficient of log F0 and in the RMSE of log F0.
As well as these simple SMs, hybrid SMs, namely SM$_{\mathrm{se}}$\&AIM and SM$_{\mathrm{es}}$\&AIM, are not promising in overall performance.

%

\subsection{Subjective evaluation test}

\begin{figure}[t]
\begin{center}
\includegraphics[width=0.9\columnwidth]{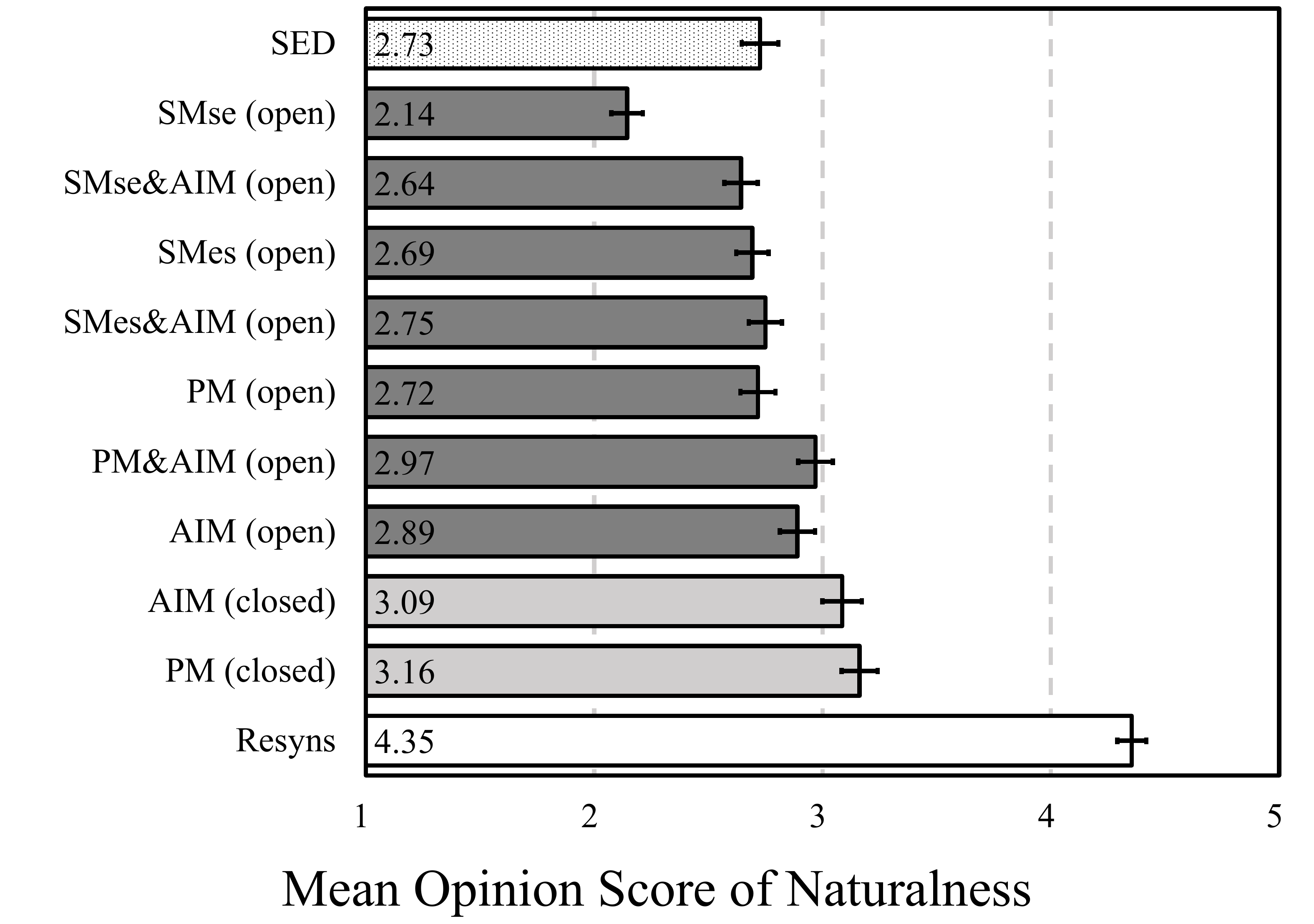}
\vspace{-15pt}     
\end{center}
\caption{MOS test of naturalness results with their 95\% confidence interval.}
\label{fig:MOS}
\begin{center}
\includegraphics[width=0.9\columnwidth]{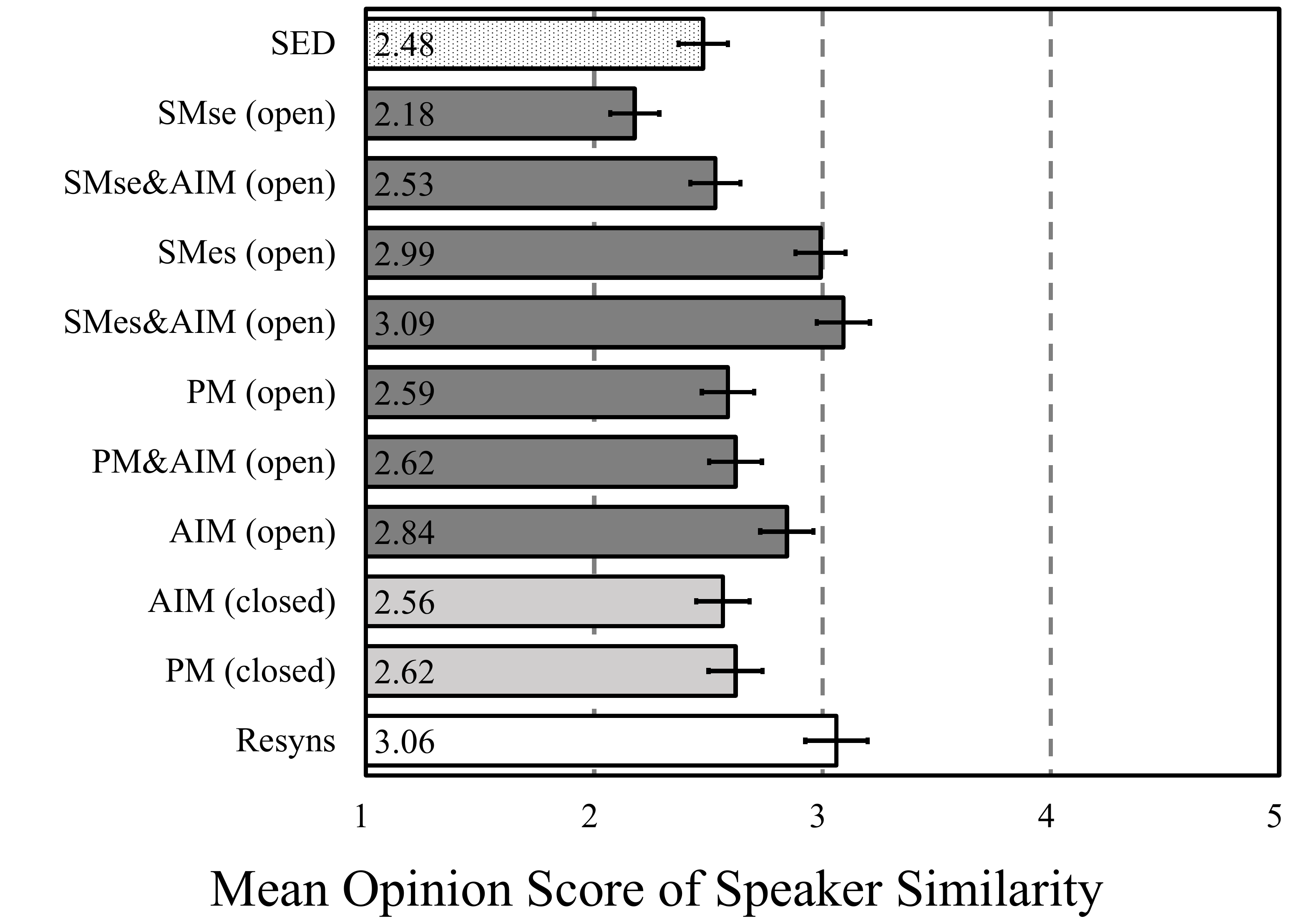}
\vspace{-15pt}     
\end{center}
\caption{MOS test of speaker similarity results with their 95\% confidence interval.}
\label{fig:DMOS}
\end{figure}

According to the results of objective evaluations, emotional expressions highly depend on each speaker. 
However, emotional expression is useful even though the means of expression is not the same as thier own means. 
Therefore, subjective tests are conducted to examine to what extent emotional expressions are reproduced in the open-emotion test.
Firstly, to confirm the basic performance, naturalness and speaker similarity are evaluated by the mean opinion score (MOS).
Then, emotion identification tests are performed in the open-emotion test.

\subsubsection{Experimental procedure of naturalness and speaker similarity}
\label{subsubsec:MOS}

To evaluate the naturalness of synthesized speech, the MOS test was carried out. 
For the open-emotion test, stimuli are synthesized by the PM, SM$_{\mathrm{se}}$, SM$_{\mathrm{es}}$, AIM, PM\&AIM, SM$_{\mathrm{se}}$\&AIM, and SM$_{\mathrm{es}}$\&AIM.
However, for the closed-emotion test, stimuli are only synthesized by the PM and AIM, because the simple and hybrid SMs showed bad performance in the correlation coefficient of log F0.
As reference speech, the SED and a resynthesized speech by STRAIGHT (resyns) are employed. 
Forty-eight sentences (twelve sentences covering two emotions from two female speakers in corpus $\alpha$ of Table \ref{tab:data_conf}) were synthesized using each model. 
A five-point scale ($1$, very unnatural, to $5$, very natural) was used for the MOS. 


To evaluate speaker similarity, the MOS test was performed, where the quality of synthesized emotional speech was compared to target neutral speech (resyns). 
The models used in the experiment were the same as the naturalness test. 
Twenty-four sentences (six sentences covering two emotions from two female speakers in corpus $\alpha$ of Table \ref{tab:data_conf}) were synthesized for each pair. 
A five-point scale ($1$, very dissimilar, to $5$, very similar) was used for the MOS. 


For both MOS tests, fifteen Japanese listeners participated. 
The order of presenting the stimuli was randomly selected, but the order was the same for all participants. 
In the speech synthesis, phoneme durations extracted from the neutral speech were used, and acoustic features were smoothed by maximum likelihood parameter generation (MLPG) \cite{bib-MLPG}. 
The variance of these features was expanded to the global variance (GV) \cite{bib-GV}, extracted from the target neutral speech, by using variance scaling \cite{bibVS}. 


\subsubsection{Experimental results  of naturalness and speaker similarity}

Figure \ref{fig:MOS} presents the results of the MOS test for naturalness. 
In the closed-emotion test, the PM and AIM show better naturalness than in the open-emotion test. 
However, the difference between closed- and open-emotion tests is smaller than the difference between Resyns and closed-emotion tests. 
In the open-emotion test, the SM$_{\mathrm{se}}$\&AIM indicates significantly better performance than SM$_{\mathrm{se}}$.
This result shows the AIM mechanism helps to improve the performance of SM$_{\mathrm{se}}$.
But this is not true for the PM and SM$_{\mathrm{es}}$.

Figure \ref{fig:DMOS} presents the results of the MOS test for speaker similarity. 
As shown in the figure, even the resynthesized voice achieves approximately $3$ in opinion score. 
This indicates that even though emotional speech is uttered by the same speaker, the speaker identity of emotional speech is a little different from that of neutral speech. 
Because of this, in terms of speaker identity, there are small differences between closed- and open-emotion tests.



\subsubsection{Experimental procedure of emotional expressions}

To evaluate the performance in extrapolating emotional expressions, emotion identification tests were carried out in the open-emotion test.
Fifteen Japanese listeners participated in the subjective test. 
For the presented synthesized speech, they were asked to select an emotion from four choices: neutral, sad, joyful, and others. 
As stimuli from the open-emotion test, the PM, SM$_{\mathrm{es}}$, AIM, PM\&AIM, and SM$_{\mathrm{es}}$\&AIM synthesized $120$ sentences (five sentences covering three emotions from four male and four female speakers in corpus $\beta$ of Table \ref{tab:data_conf}), the total number of sentences was $600$. 
As a reference stimuli from the closed-emotion test, the PM is selected.
From informal listening tests, the SM was worse, but there are no significant differences between the AIM and PM.
Besides, the PM has averaged performance for objective evaluations.
The PM synthesized $30$ sentences (five sentences covering three emotion from two female speakers in corpus $\alpha$ of Table \ref{tab:data_conf}).
The order of presenting the stimuli was randomly selected, but the order was the same for all participants. 
The speech synthesis procedures are the same as \ref{subsubsec:MOS}.
A chi-square test was used for evaluating whether the open-emotion test has a significant difference with the closed-emotion test in each correct emotion. 

\subsubsection{Experimental results of emotional expressions}

Table \ref{tab:ex4_acc} presents the confusion matrices of participants' choices and correct answers.
A symbol (*) indicates p$<0.001$ in a chi-square test between the closed-emotion by the PM (f) and the others (a, b, c, d, and e).
For neutral, all models have a high correct rate ($>0.85$). 
This finding is reasonable because neutral is the closed-emotion test in all models. 
For sad, little difference is observed between the closed-emotion test (Table \ref{tab:ex4_acc} (f)) and open-emotion test (Table \ref{tab:ex4_acc} (a), (b), (c), (d), and (e)). 
This mainly occurs because the differences in F0 and the cepstrum for sad are relatively small in Fig. \ref{fig:logF0_dist} and Fig. \ref{fig:mcep_dist}, and easily trained from other speakers' speech. 
Based on the results, we propose that sadness can be expressed by the proposed models (PM, SM$_{\mathrm{es}}$, AIM, PM\&AIM, and SM$_{\mathrm{es}}$\&AIM). 
For joyful, however, only the PM demonstrates little difference in the closed-emotion test and open-emotion test. 
In Fig. \ref{fig:logF0_dist} and Fig. \ref{fig:mcep_dist}, differences in F0 and the cepstrum for joyful are larger than those for sad.
The results indicate that the PM can independently model the speaker factor and emotional factors even if large differences are observed in the acoustic parameters to express emotion. 
According to the results, we should select the PM from the proposed five models to synthesize emotional speech based on the extrapolation approach.

\section{Conclusion and future work}\label{sec:Conclusion}
In this paper, to generate emotional expressions using DNN-based TTS, we proposed the following five models: PM, SM, AIM, PM\&AIM, and SM\&AIM. 
These models are based on the following extrapolation approach: emotional expression models are trained using speech uttered by a particular person, and the models are applied to another person for generating emotional speech with the person's voice quality. 
In other words, the collection of emotional speech uttered by a target speaker is unnecessary, and emotional expression is generated using models trained by the emotional speech of another person. 
To evaluate the extrapolation performance of the proposed models, an open-emotion test and closed-emotion test were conducted objectively and subjectively. 
The objective evaluation results demonstrate that the performances in the open-emotion test are insufficient on the basis of a comparison to those in the closed-emotion test, because each speaker has their own manners of expressing emotion. 
However, the subjective evaluation results indicate that the proposed models can convey emotional information to some extent, especially, the PM, which can correctly convey sad and joyful emotions at a rate of ${>}60$\%. 
In conclusion, the PM architecture works well to separately train the speaker and emotion factors for DNN-based TTS.

The proposed methods synthesize emotional speech possible without collecting emotional speech uttered by a target speaker. 
However, in evaluation experiments, each speaker has their own manners of expressing emotion. 
The solution to this problem is a topic for further research.
First, the needs should be examined, that is, in which situation speaker-specific emotional expressions are necessary and how precisely the emotional expression should be reproduced. 
The other consideration is a trade-off between collecting emotional speech uttered by target speakers and the performance of the extrapolation approach. 
As mentioned in Section 1 (Introduction), it is difficult for individuals to utter speech with a specific emotion, and to continue speaking with that emotion for an extended time. 
Thus, no guarantee can be given that models trained by collecting emotional speech uttered by target speakers always outperform the performance of the extrapolation approaches.
The performance of the DNN-based extrapolation method will be clarified by comparing it with conventional extrapolation methods.

%
%
%

%
%



\bibliographystyle{IEEEbib}

\bibliography{k_inoue_journal}

\end{document}